\documentclass[%
 reprint,
superscriptaddress,
 amsmath,amssymb,
 aps,
]{revtex4-1}

\usepackage{graphicx}
\usepackage{dcolumn}
\usepackage{bm}
\usepackage{color}	
\usepackage{hyperref}

\newcommand{\mps}[0]{m$\cdot$s$^{-1}$}

\begin{document}
\title{Simultaneous concentration and velocity maps in particle suspensions under shear from rheo-ultrasonic imaging}
\author{Brice Saint-Michel}
\affiliation{Univ Lyon, Ens de Lyon, Univ Claude Bernard, CNRS, Laboratoire de Physique, F-69342 Lyon, France}
\email{brice.saint-michel@ens-lyon.fr}

\author{Hugues Bodiguel}%
\affiliation{Univ Grenoble Alpes, Univ Joseph Fourier, CNRS, INP Grenoble, Laboratoire Rh\'eologie et Proc\'ed\'es, F-38041 Grenoble, France}%

\author{Steven Meeker}
\affiliation{Univ Bordeaux, Univ Bordeaux 1, CNRS, Solvay, Laboratory of the Future, F-33608 Pessac, France}%

\author{S\'ebastien Manneville}
\affiliation{Univ Lyon, Ens de Lyon, Univ Claude Bernard, CNRS, Laboratoire de Physique, F-69342 Lyon, France}%

\date{\today}

\begin{abstract}
We extend a previously developed ultrafast ultrasonic technique [Gallot {\it et al.}, {\it Rev. Sci. Instrum.} {\bf 84}, 045107 (2013)] to concentration field measurements in non-Brownian particle suspensions under shear. The technique provides access to time-resolved concentration maps within the gap of a Taylor-Couette cell simultaneously to local velocity measurements and standard rheological characterization. Benchmark experiments in homogeneous particle suspensions are used to calibrate the system. We then image heterogeneous concentration fields that result from centrifugation effects, from the classical Taylor-Couette instability and from sedimentation or shear-induced resuspension.
\end{abstract}

\pacs{Valid PACS appear here}
\maketitle

\section{Introduction}
Non-Brownian suspensions, where particles typically ranging in size from a few micrometers to over $100~\mu$m are dispersed in a suspending fluid, encompass a wide range of natural and man-made materials.  These include suspensions as diverse as mud, blood, paints, ice creams, and flour preparations, to name but a few. The flow behavior of such mixtures is known to raise both important practical difficulties and theoretical challenges~\cite{Mewis2012}. In the case of hard-sphere-like particles, one of the most prominent results is the transition from a shear-thinning regime in dilute and semi-dilute suspensions to a strong shear-thickening state for highly concentrated suspensions~\cite{Metzner1958}. From the practical point of view, shear thickening often limits the processesing of industrial suspensions. It is also widely known that particles may migrate in suspensions under flow due to a large array of physico-chemical processes including gravity, pressure and shear gradients, hydrodynamic interactions, electric potentials and magnetic fields. Particle migration in turn affects the rheology of the suspension itself. Indeed, it has been recently invoked as a key factor at play in discontinuous shear thickening~\cite{Fall2015}. Shear-induced resuspension~\cite{Leighton1986} is a classical example where a flow and a heterogeneous particle concentration field act on each other in a suspension. Such an interplay between flow and particle volume fraction is of great significance for various applications, e.g. solid-liquid separation in wastewater treatment of industrial slurries and in the food, cosmetics and pharmaceutical industries~\cite{Landman:1994,Boger:2000}.

To fully characterize  the dynamics of flowing particle suspensions  requires experimental techniques that give access to both local concentration and velocity fields, especially when particle migration leads to inhomogeneous  suspension flows. Furthermore, when such techniques can be used simultaneously for rheological characterization, they provide crucial insights into the coupling between particle dynamics and the macroscopic flow behavior of the suspension. For transparent suspending fluids and large enough particles, as in the case of geophysical flows, direct particle tracking is possible~\cite{Houssais:2015}. However, suspensions of smaller particles are generally opaque for volume fractions above a few percent, so the optical investigation of denser systems generally involves more advanced techniques such as confocal scanning light microscopy \cite{Manneville:2008,Besseling:2009} or depolarized light scattering \cite{Brambilla:2011}. Moreover, light-based techniques can hardly be implemented for strongly light-scattering suspending fluids (e.g. milk). Magnetic resonance imaging offers an interesting alternative as it is accurate at any particle volume fraction whatever the turbidity of the suspending fluid \cite{Callaghan:1999,Rodts:2010}. Still, this costly technique is tricky to couple with standard rheological setups and generally requires long integration times, limiting its use to slowly evolving systems. 

In this paper, we demonstrate the feasibility of simultaneous rheology, velocimetry and local volume fraction measurement from ultrafast ultrasonic imaging. We extend a recently developed rheo-ultrasonic setup~\cite{Gallot2013} to access the concentration field in non-Brownian suspensions subject to both homogeneous and heterogeneous flows. In Section~\ref{sec:AcousInt}, we briefly review previous work on acoustic characterization of suspensions and recall the technical specifications of our ultrasonic scanner. We then detail our method for computing the local ultrasonic amplitude $P$ across the gap of a Taylor-Couette device. Section~\ref{sec:calib} is devoted to the determination of the local volume fraction $\phi(\mathbf{r})$ from the acoustic amplitude maps $P(\mathbf{r})$ and to the calibration of the system in homogeneous suspensions of particles made of various materials. Finally, in Section~\ref{sec:Examples}, we illustrate the possibilities of the technique on several heterogeneous flows in dilute and semi-dilute suspensions.

\section{Acoustic imaging in suspensions of non-Brownian particles}
\label{sec:AcousInt}

\subsection{A brief review of acoustic characterization of suspensions}
\label{sec:review}

Suspensions of solid particles or of liquid droplets have been studied using acoustic techniques for more than two decades. A widely used technique is ultrasonic spectroscopy which consists in measuring the sound speed and the attenuation of ultrasound as a function of frequency \cite{Dukhin:1996a,Dukhin:1996b,Povey:1997}. Provided a model is assumed and/or some calibration step is performed, this technique leads to an estimate of the global suspension volume fraction $\phi_0$ and even to the determination of some physical characteristics of the particles such as their size, their polydispersity or their elastic modulus \cite{Challis:2005,Thorne:2014}. Applications of such acoustic measurements include the characterization of food products \cite{McClements:1997,Awad:2012}, emulsions \cite{McClements:1989}, or more recently clay suspensions \cite{Ali:2013}.

Since it relies on the continuous transmission of a harmonic ultrasonic wave, the information is integrated from the emitter to the receiver. Ultrasonic spectroscopy is thus only a \textit{global} technique. In order to recover \textit{local} information from ultrasound, one has to turn to ``pulsed'' techniques where short bursts of ultrasonic waves are sent through the suspension. When these pulses are reflected by the particles, they give rise to a collection of short echoes that can be recorded as a function of time by a receiver or by an array of receivers. If one can neglect the contributions of subsequent reflections of the singly-scattered waves by different particles, i.e. if multiple scattering can be neglected, the arrival times $t$ of these echoes directly relate to the spatial position $r$ (or ``depth'') of the particles along the acoustic beam through $r=ct/2$ where $c$ is the sound speed of the suspension and the factor 1/2 accounts for the round-trip propagation. This time-space correspondence constitutes the basic principle of ultrasonic echography and the reader is referred to Ref.~\cite{Szabo:2004} for general technical information on the subject.

 \begin{figure}
 	\centering
 	\includegraphics{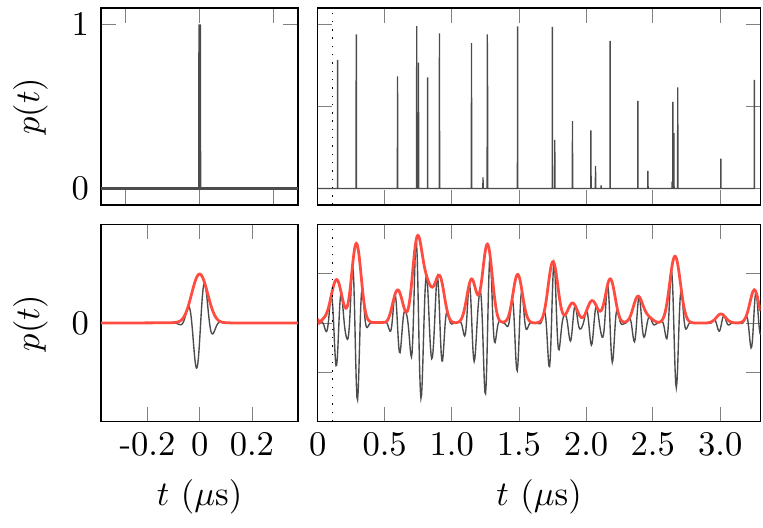}
     \caption{(top panels) Emitted ultrasonic pressure signal $p(t)$ in the ideal case of an infinitely short pulse (left) and backscattered signal received by one transducer of the echographic scanner (right). (bottom panels) Realistic case where the emitted pulse (left) has a central frequency of 15~MHz and a finite width. The backscattered signal exhibits a speckle pattern with positive and negative lobes (black line). An estimate of the local concentration can be computed by using the Hilbert transform of the speckle signal (thick red line) as long as multiple scattering remains negligible so that the time-space correspondence $r=ct/2$ is justified.}
     \label{fig:THilb}
 \end{figure}

In the case of very dilute suspensions, individual particles can be counted and tracked from echographic images. If infinitely short ultrasonic pulses with negligible beam width could be sent and recorded as illustrated in the top panels of Fig.~\ref{fig:THilb}, such tracking would remain possible whatever the volume fraction. However, in practice, the acoustic beam has a finite width which may encompass more than one scatterer at a given depth. Moreover, the pressure pulse itself has a finite temporal length and bandwidth. Therefore, when the volume fraction increases, reflections from the particles are no longer well separated in the temporal ultrasonic signals $p(t)$: they rather form an ``ultrasonic speckle'' that results from the interference of the various waves scattered by the particles, as sketched in the bottom panels of Fig.~\ref{fig:THilb}. This speckle can still be used to assess heterogeneities in a material, e.g. biological tissues in diagnostic imaging \cite{Szabo:2004}, or to follow the particle displacement in a flow through Ultrasound Imaging Velocimetry (UIV) \cite{Poelma:2012}, an acoustic equivalent of optical Particle Imaging Velocimetry (PIV) also known as Ultrasonic Speckle Velocimetry (USV) \cite{Sandrin:2001,Manneville:2004a} or echo-PIV \cite{Kim:2004}. UIV and its derivatives have proved successful not only for blood flow measurements \cite{Zhang:2011,Tanter:2014} but also for applications to hydrodynamics \cite{Takeda:1999}, to rheology \cite{Gallot2013,Gurung:2016} or to industrial situations involving optically opaque suspensions \cite{Wiklund:2010,Kotze:2013}. These velocimetry techniques rely upon the measurement of local displacements through cross-correlation of successive speckle signals received after successive pulses are sent through the scattering material. Such data analysis is generally rather straightforward to implement thanks to standard post-processing algorithms. It easily provides robust velocity estimates as it is based on phase shift measurements which are not very sensitive to the signal intensity (as long as this intensity is large enough).

Analyzing the speckle intensity in order to recover quantitative local information on the particle volume fraction or on other material heterogeneities turns out to be more difficult. In the specific context of marine suspended sediments \cite{Young1982,Hanes1986}, the following semi-empirical relation has been established between the local volume fraction $\phi(r)$ and the root mean square backscattered pressure signal $p_{\rm rms} (r)$ for large dilutions $\phi_0\lesssim0.01$ and in the far-field of a plane acoustic transducer~\cite{Thorne1991,Thorne1993}:
\begin{equation}
	\label{eq:prms}
	p_{\rm rms} (r) = \frac{K_1}{r} \phi(r)^{1/2} \exp \left [-\alpha(r) r \right ]\,,
\end{equation}
where $p_{\rm rms}(r)$ is computed from an ensemble-average of the backscattered signal $p(t)$ over several successive pulses --and thus over statistically different configurations of the particles-- and by using the time-space correspondence $r=ct/2$. In Eq.~(\ref{eq:prms}),  $K_1$ is a proportionality factor with dimension of Pa$\cdot$m, the factor $1/r$ corresponds to the far-field divergence of the acoustic beam, and $\alpha(r)$ accounts for the attenuation of the ultrasound over the whole round trip from the transducer to distance $r$ and back to the transducer. If attenuation can be neglected [i.e. if $\alpha(r)=0$], Eq.~(\ref{eq:prms}) indicates that $p_{\rm rms}(r)$ is directly proportional to the square root of the local volume fraction, which simply expresses the fact that the backscattered waves contribute incoherently to $p_{\rm rms}(r)$. More generally the ultrasonic attenuation can be divided into two terms,  $\alpha(r) = \alpha_f + \alpha_p(r)$, where $\alpha_f$ corresponds to the absorption coefficient of the ultrasound by the suspending fluid and $\alpha_p(r)$ is the additional attenuation induced by the presence of suspended particles that ``shade'' the particles behind them due to scattering. This second source of attenuation is space-dependent and within a single-scattering approximation, it can be written as~\cite{Thorne1991,Thorne1993}:
\begin{equation}
	\label{eq:atten}
    \alpha_p(r) = \frac{K_2}{r} \int_{r_0}^r  \phi(u) {\rm d} u\,,
\end{equation}
where $r_0$ corresponds to the position at which the ultrasonic beam enters the suspension and $K_2$ is a an intrinsic attenuation coefficient analogue to the molar attenuation coefficient of the Beer-Lambert law. Like $K_1$ in Eq.~(\ref{eq:prms}), $K_2$ depends on the fluid properties, on the type of particles and on temperature in a complex manner.

The fact that the integral in Eq.~(\ref{eq:atten}) involves the local volume fraction itself seriously complicates the inversion of Eq.~(\ref{eq:prms}) in order to recover $\phi(r)$ in the general case of a heterogeneous suspension. Moreover, when the particle volume fraction in the suspension increases typically above a few percents, multiple scattering becomes predominant so that the time-space correspondence required for imaging the suspension is lost \cite{Page:1995,Page:2000,Tourin:2000}. Besides, Eq.~(\ref{eq:prms}) is no longer valid if significant multiple scattering occurs. These difficulties probably explain why applications of ultrasonic imaging to study the local particle concentration in suspensions have remained very scarce until recent years \cite{Hunter:2012,Zou:2014,Bux:2016}. In the following, we shall build upon a previous rheo-ultrasonic setup designed for local velocimetry of complex fluid flows within a concentric cylinder (Taylor-Couette, TC) geometry \cite{Gallot2013} to perform local volume fraction measurements. Multiple scattering is minimized by using a small-gap TC cell. By playing on the physical properties of the particles, we show that large volume fractions can be accessed, up to global volume fractions $\phi_0\simeq 0.4$ in the most favorable case. Frame rates up to 20,000~fps are achieved thanks to plane wave imaging.

\subsection{Ultrasonic imaging coupled to rheometry}

 \begin{figure}
 	\centering
 	\includegraphics[scale=0.60]{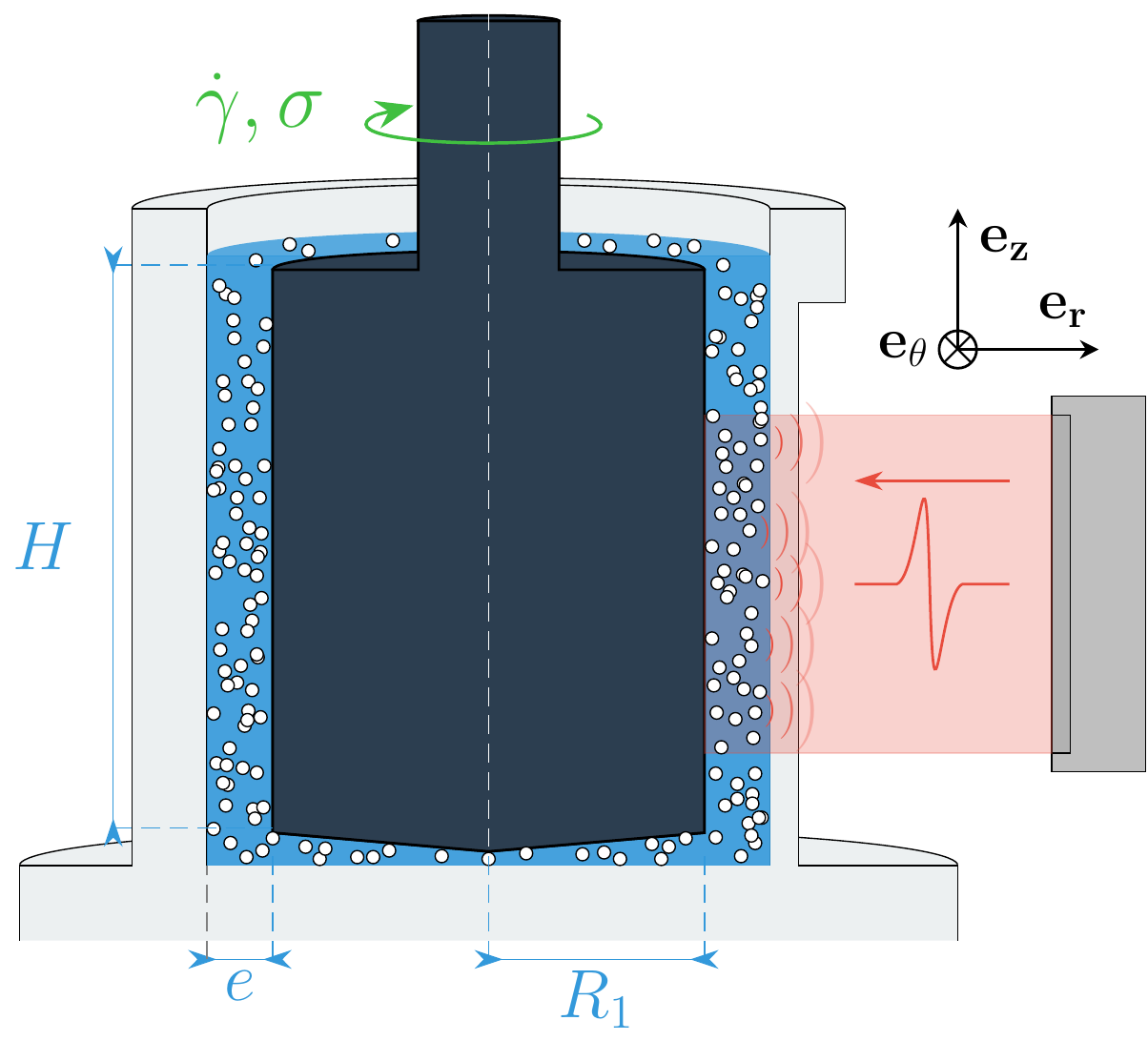}
     \caption{Illustration of the experimental setup and notations. The suspension is confined within the gap $e$ between two concentric cylinders, the rotor of radius $R_1$ and height $H$ (in black) and the stator of inner radius $R_2=R_1+e$ (in grey). The rotation of the rotor is controlled by a stress-imposed rheometer (not shown) that measures the shear rate $\dot\gamma$ and the shear stress $\sigma$ imposed to the suspension. The stator is immersed in a water tank and an ultrasonic array (in dark grey) emits plane ultrasonic pulses (in red) and records the echoes backscattered by the suspended particles.}
     \label{fig:setup}
 \end{figure}

The experimental setup used in the present study has already been described at length in Gallot~\textit{et al.}~\cite{Gallot2013} and is illustrated in Fig.~\ref{fig:setup}. It consists of a Taylor-Couette shear cell made of a static cylindrical cup of inner radius $R_{2} = 25$~mm, hereafter referred to as the \emph{stator}, and a smaller cylinder of radius $R_{1} = 23$~mm and height $H = 60$~mm, called the \emph{rotor}, whose rotation is driven by a standard stress-imposed rheometer (TA Instruments ARG2). The cylinder axis is aligned with the vertical axis ${\bf e_{\rm z}}$, the radial and azimuthal directions following the classical ${\bf e_{\rm r}}$ and ${\bf e_{\theta}}$ unit vectors. The gap between the two cylinders is thus $e=R_2-R_1=2$~mm. Temperature regulation at $T=25^\circ$C is achieved using a water bath surrounding the TC cell. The cell is filled with different particle suspensions, as detailed below in Sect.~\ref{ssec:samples} and as listed in Table~\ref{tab:props}.  Thanks to a feedback loop on the torque applied to the rotor, this type of rheometer can impose a constant shear rate $\dot\gamma$ on the sample. 

\begin{figure*}
	\centering
	\includegraphics[scale=0.98]{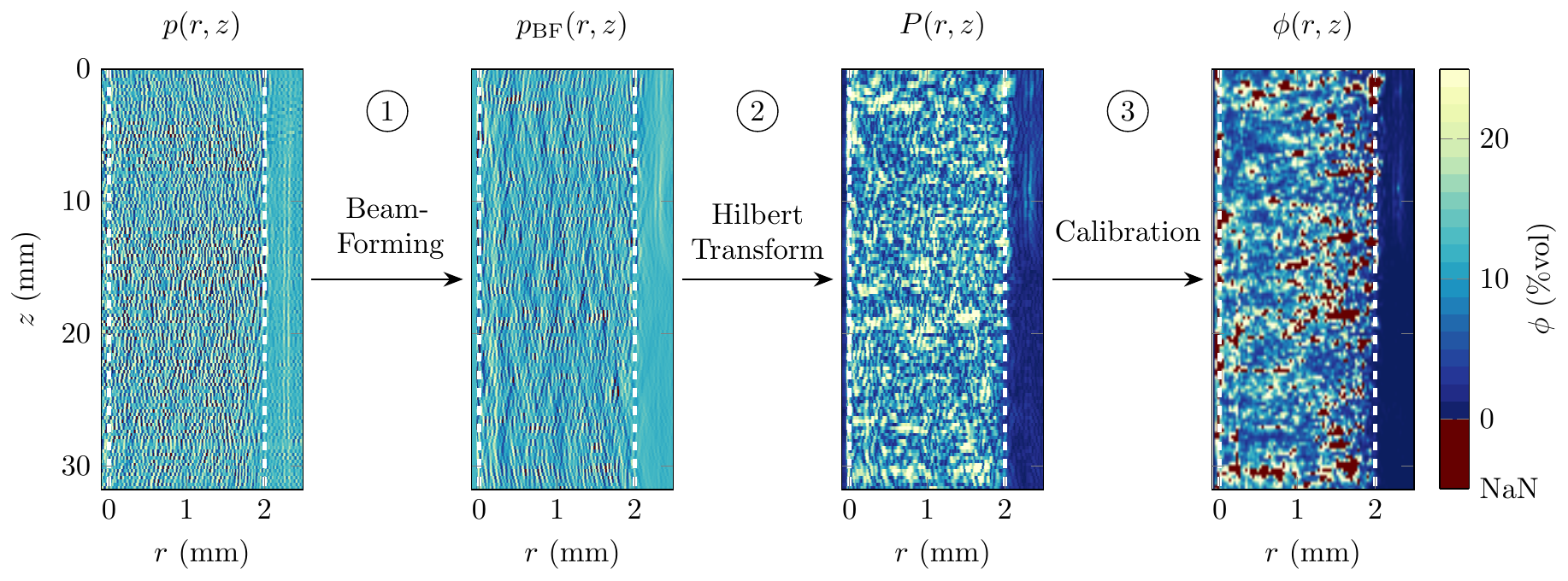}
    \caption{The three successive data processing steps involved for measuring local volume fractions in suspensions sheared in a Taylor-Couette device. Step~1 is a beam-forming step, described in details in Ref.~\cite{Gallot2013}, that allows one to go from the raw ultrasonic speckle signal $p(r,z)$ to the beam-formed image $p_{\rm BF}(r,z)$. Step $2$ consists of a signal envelope detection through a Hilbert transform (see Section~\ref{sec:Hilbert}). Step $3$ performs a sliding average over successive individual Hilbert maps $P(r,z)$ and computes the local volume fraction $\phi(r,z)$ through table lookup in calibration plots as described in Section~\ref{sec:prmdet}. In all the images, the abscissa $r$ denotes the radial distance from the rotor and the ordinate $z$ corresponds to the vertical distance with $z=16$~mm taken roughly at the middle height of the Taylor-Couette cell. Dashed lines show the positions of the rotor ($r=0$) and of the stator ($r=e=2$~mm). In the rightmost image, the average is taken over 10 successive maps and dark red codes for the regions where the calibration cannot be performed unambiguously and that are therefore associated with ``Not a Number'' (NaN) values (see text). Data obtained in a suspension of polystyrene spheres of diameter 20~$\mu$m in water with a volume fraction $\phi_0=0.15$ and sheared at $\dot\gamma=20$~s$^{-1}$.}
    \label{fig:Process_Phi}
\end{figure*}

The sample is insonified with ultrasonic waves emitted by a linear array of 128 piezoelectric transducers stacked over 32~mm along the vertical direction and controlled by an ultrasonic scanner (Lecoeur Electronique Open S). The transducer array is immersed in the water bath which serves as an acoustic transmission medium. All transducers simultaneously emit a short pulse with central frequency $15$~MHz and typical length of 3 acoustic periods, thus generating a pulsed plane wave that propagates along a vertical plane and crosses the sample at a given angle $\theta$ relative to the radial direction. This pulsed wave is scattered by the suspended particles and the corresponding backscattered speckle signal $p(r,z)$ is recorded by each transducer and stored in the scanner memory for post-processing (see leftmost panel in Fig.~\ref{fig:Process_Phi} for a typical speckle signal). Our electronic device can send up to 20,000 pulsed waves per second, allowing ultrafast imaging of the sample \cite{Tanter:2014}. 

As detailed in Ref.~\cite{Gallot2013}, further data processing of the speckle signal involves the removal of fixed echoes using a reference average speckle, followed by a \emph{parallel beam-forming} step to deconvolve the signal of each channel. This first step, indicated as step 1 in Fig.~\ref{fig:Process_Phi}, leads to beam-formed images $p_{\rm BF} (r,z)$ which can then be used in correlation-based algorithms to retrieve the velocity field component $v_{\rm \theta}(r,z)$ across the (${\bf e_r}, {\bf e_z}$) plane. Here $r$ denotes the radial distance from the rotor so that $r=e=2$~mm corresponds to the position of the stator (see dashed lines in Fig.~\ref{fig:Process_Phi}). $r$ is deduced from the speckle arrival time using the space-time correspondence already mentioned above and from prior determination of the positions of the cell walls and of the angle of incidence $\theta$ based on a calibration of velocity profiles measured in a very dilute ($\phi_0=0.01$), Newtonian suspension of the test particles in the suspending fluid. This calibration requires one to measure the speed of sound $c$ independently both in the suspending fluid (see Table~\ref{tab:props}) and in the suspension. Although $c$ is known to increase with volume fraction \cite{Stolojanu:2001}, it changes by at most 5~\% for volume fractions up to 0.35, which leads to negligible corrections on the positions of the cell walls or on $\theta$. Therefore, for a given suspending fluid and for a given type of particles, these parameters are kept constant and equal to those found for the very dilute case. The reader is referred to Ref.~\cite{Gallot2013} for full technical details on step 1. The vertical resolution of the images is given by the pitch of the transducer array, $\delta z=0.25$~mm, the longitudinal resolution $\delta r$ corresponds to the acoustic wavelength $\lambda=0.1$~mm, and the width of the ultrasonic beam within the gap of the TC cell is about $0.7$~mm.

\subsection{Processing of the speckle amplitude}
\label{sec:Hilbert}

Our previous work~\cite{Gallot2013} was restricted to using the \emph{phase} information of the beam-formed images $p_{\rm BF}(r,z)$ through cross-correlation of successive images in order to recover the speckle displacement and hence the fluid velocity field. Here, our goal is to extend the technique by also processing the \emph{amplitude} information conveyed by the backscattered signal. Indeed, as explained above in Sect.~\ref{sec:review}, the amplitude of $p_{\rm BF}(r,z)$ carries key information on the local volume fraction. Rather than focusing on $p_{\rm rms}$, we compute the local speckle amplitude $P(r,z)$ of every beam-formed image by using the Hilbert transform $H(p_{\rm BF})$ of $p_{\rm BF}$ along the variable $r$:
\begin{align}
	\label{eq:Thilb}
	P(r,z) &= \left | p_{\rm BF}(r,z)+ iH(p_{\rm BF})(r,z) \right | \\
	&= \left | p_{\rm BF}(r,z) + \frac{i}{\pi} \int_{-\infty}^{\infty} \frac{p_{\rm BF}(s,z)}{r -s} {\rm d}s \right |\,.
\end{align}
This constitutes step 2 in Fig.~\ref{fig:Process_Phi} and the resulting signal $P(r,z)$ captures the envelope of the original oscillating signal, which to some extent reflects the amount of particles present locally in the TC cell.

$P(r,z)$ is further averaged over several successive pulses and used as an input for the determination of the local volume fraction $\phi(r,z)$. This last step is based on calibration experiments detailed below in Sect.~\ref{sec:calib}.

\section{From ultrasonic speckle amplitude to local volume fraction}
\label{sec:calib}

In this section, we quantitatively link the local ultrasonic amplitude $P(r,z)$ to the volume fraction in various homogeneous suspensions. This allows us to build a calibration database for specific suspending fluids and particles, before turning to the investigation of spatially heterogeneous particle distributions in Sect.~\ref{sec:Examples}.

\subsection{Samples}
\label{ssec:samples}

In order to explore the influence of particle size and material, we focus on five different types of spherical particles: polystyrene spheres (Microbeads Dynoseeds TS-20 and TS-80, respectively referred to as ``PS-1'' and ``PS-2''), polymethylmethacrylate spheres (Microbeads Spheromers CA-40, hereafter noted ``PMMA''), glass spheres (Cospheric P2075SL), and polyamide spheres (Arkema Orgasol 2002 ES3 NAT3, noted ``PA''). These particles show limited polydispersity so that we shall consider them as monodisperse in the following. 

Since the physical properties of the suspending fluid also affect ultrasonic attenuation and scattering, we test three different fluids: distilled water, a 27~vol.~\% water--73~vol.~\% glycerol mixture (noted ``H$_2$O-Gly''), and a light mineral oil (Sigma Aldrich, CAS number: 8042-47-5). All particles could be easily dispersed in water or in water-glycerol mixtures except for the PA particles. Experiments on PA particles were therefore performed in light mineral oil. The physical characteristics of the various suspension constituents are gathered in Table~\ref{tab:props}.

\begin{table}[h]
	\begin{tabular}{l c c c c c c}
    \hline
    			  & $\nu$			   & 		$a$ & $\Delta a/a$ 		& $\rho$ 				& $c$  			    & $Z$ \rule{0pt}{12pt}  \\
    	 		     &   {\scriptsize $10^{-6}$~m$^2\cdot$s$^{-1}$}  &  {\scriptsize $\mu$m}	&  & {\scriptsize kg\,m$^{-3}$}		&  {\scriptsize m\,s$^{-1}$}  & {\scriptsize $10^{6}$~kg\,m$^{-2}$\,s$^{-1}$}	\rule{0pt}{12pt}    \\
    \hline
    	 	H$_2$O	 & 	0.90			  &		&			& 1000					& 1500			   	& 1.50\rule{0pt}{12pt}    \\
	H$_2$O-Gly	 &		29	  &					&		& 1200					& 1850				& 2.22			  		  \\
    	 		Oil		 &		26		  &			&			& 830					& 1460				& 1.21				      \\
     \hline
     	 	PS-1		 &		  &				20		&	0.1	& 1050					& 2340				& 2.46\rule{0pt}{12pt}    \\
        	PMMA		 &  		  &		40		& 0.05			& 1200					& 2700				& 3.24					\\
 	Glass		 &  &						70		& 0.5		& 2540					& 5340				& 15.0					\\
        	PS-2		 &		  &			80		& 0.1			& 1050					& 2340				& 2.46					\\
         	PA			 &  	      &		30		& 0.3			& 1030					& 2200				& 2.27					\\
 	\hline
	\end{tabular}
    \caption{Physical properties relevant to ultrasonic attenuation and scattering of the suspension constituents at 25$^\circ$C: kinematic viscosity $\nu$ of the suspending fluids and diameter $a$ of the particles together with an estimate of their polydispersity $\Delta a/a$ as indicated by the manufacturer, their density $\rho$, sound speed $c$ and acoustic impedance $Z=\rho c$. Data taken from the CRC Handbook of Chemistry and from the olympus.com website.}
    \label{tab:props}
\end{table}

\subsection{Benchmark experiments in homogeneous suspensions}
\label{sec:benchmark}

\begin{figure*}
	\centering
    \includegraphics{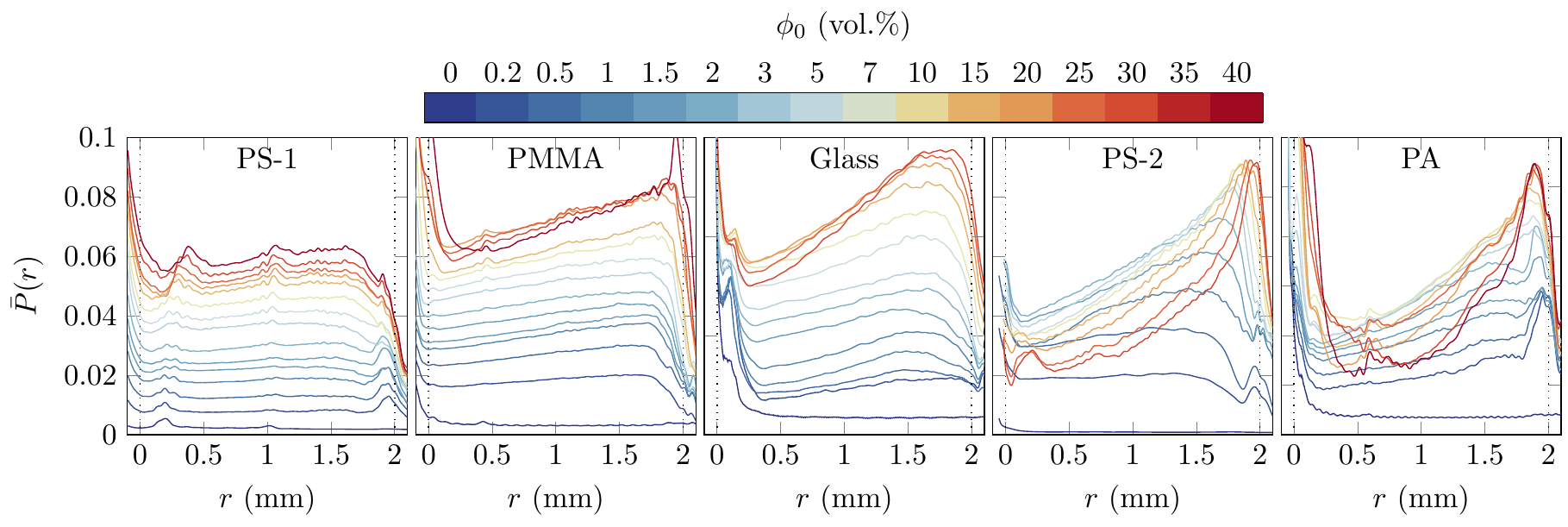}
    \caption{Benchmark experiments in various homogeneous suspensions with global volume fraction $\phi_0$. The graphs show the time- and $z$-averaged ultrasonic amplitude $\bar{P}(r)$ for five different types of particles suspended in different fluids as listed in Table~\ref{tab:props}. The amplitude level was scaled according to the specific transducer gain used for each particle type. The incident ultrasonic pulses propagate from the stator at $r=e=2$~mm to the rotor at $r=0$ (see dashed lines). From left to right: polystyrene spheres (PS-1, diameter $a = 20~\mu$m) in water, PMMA spheres ($a = 40~\mu$m) in water, glass spheres ($a = 70~\mu$m) in a 27~\% vol. water--73~\% vol. glycerol mixture, polystyrene spheres (PS-2, $a = 80~\mu$m) in water, and polyamide spheres (PA, $a = 30~\mu$m) in mineral oil. The color codes for the volume fraction $\phi_0$.}
    \label{fig:C_r}
\end{figure*}

Benchmark experiments for the five different particle types listed in Table~\ref{tab:props} are presented in Fig.~\ref{fig:C_r}, which plot the time- and $z$-averaged ultrasonic amplitude $\bar{P}(r)$ of the suspensions across the Couette gap. In these experiments, for a given global volume fraction $\phi_0$, the suspensions are homogenized by first raising the rotor so that its bottom matches the fluid surface level and then rotating the rotor at a high shear rate (typically $\dot\gamma = 10^3$~s$^{-1}$). This produces a strong, turbulent mixing flow that is maintained for about 30~s, after which the rotor is lowered to its usual position, i.e. down to 100~$\mu$m from the stator bottom, while rotating at a constant angular velocity corresponding to the target shear rate $\dot\gamma = 20$~s$^{-1}$. This shear rate is large enough to limit sedimentation through flow-induced resuspension but low enough to ensure that the flow remains laminar and purely tangential in all cases. 

Reference experiments are first performed in the absence of particles, i.e. for $\phi_0=0$, and the corresponding average speckle data is then subtracted to all subsequent speckle images. As explained in Ref.~\cite{Gallot2013}, this procedure allows us to remove spurious fixed echoes that arise from reflections on the cell walls, on the bottom of the water tank and at the water surface, provided the sound speed remains constant in the material. The efficiency of this procedure and the level of electronic noise on speckle data can be assessed thanks to independent experiments at $\phi_0=0$ (see the bottommost purple lines in Fig.~\ref{fig:C_r}). The particle volume fraction is then systematically increased from $\phi_0=0.005$ up to $\phi_0=0.4$ by progressively adding weighted amounts of particles and mixing the suspensions again. 

Ultrasound data acquisition starts with a pulse repetition frequency of 1000 pulses per second as soon as the imposed shear rate $\dot\gamma = 20$~s$^{-1}$ is reached. The ultrasonic amplitude $P(r,z)$ is averaged over 1000 different speckle configurations. This corresponds to an acquisition time of 1~s, over which possible migration effects at the cell boundaries can be neglected. Since sedimentation is also negligible, we further average $P(r,z)$ over the vertical direction $z$. Still, due to differences in particle size and acoustic impedance and/or in transducer gain, the absolute level of ultrasonic amplitude is not comparable from one type of suspension to the other. For a given particle type and suspending fluid, the transducer receiving gain is therefore adjusted to offer both a good sensitivity at low concentrations and a unsaturated speckle signal at high concentrations. This provides the time- and $z$-averaged amplitude profiles $\bar{P}(r)$ shown in Fig.~\ref{fig:C_r}.

As expected, the general behavior of $\bar{P}(r)$ with both $r$ and $\phi_0$ strongly depends on the particular suspension. We recall that $r=0$ corresponds to the rotor and that $r=e=2$~mm corresponds to the stator so that ultrasonic pulses propagate from the right to the left in the plots of Fig.~\ref{fig:C_r}. In all cases, both walls are clearly identified. At the stator, the amplitude sharply decreases for $r\gtrsim 1.9$ and the stator is sometimes seen as a small maximum in $\bar{P}(r\simeq e)$ (see PS-1 and PS-2 at low concentrations). On the rotor side, a large increase of the amplitude is observed for $r\lesssim 0$--0.1~mm, which signals the presence of the rotor whose reflection can never be perfectly cancelled out with the procedure described above.

We also note that the region where $\bar{P}(r)$ increases close to the rotor tends to shift to larger values of $r$ as the concentration is increased, typically above $\phi_0\gtrsim 0.2$ (see PMMA and PA). This may be interpreted as the consequence of the increase of the sound speed with $\phi_0$, which leads to a smaller apparent gap size. Still, the fact that the rotor can be detected from speckle images is an indication that multiple scattering remains limited thanks to the use of a small-gap TC cell (see also discussion below in Sect.~\ref{sec:physlim}). To further check for multiple scattering and for the influence of the rotor on our measurements, we  performed ultrasonic imaging in the absence of the rotor. The results, shown in Supplemental Figure~1, allow us to conclude that, up to systematic artifacts that are probably due to fixed echoes that were not properly removed (see $r\simeq 0.25$ for PS-2 and $r\simeq 0.5$ for PA), we can confidently interpret the evolution of $\bar{P}(r)$ with $r$ and $\phi_0$ as the signature of the local volume fraction $\phi(r)$ up to within 0.1~mm from the TC cell boundaries.

\subsection{Calibration of local volume fraction measurements}
\label{sec:prmdet}

For all particle types, the ultrasonic amplitude $\bar{P}(r)$ is generally seen to decrease for decreasing $r$. This is a direct signature of the attenuation of the acoustic wave within the suspension. More precisely, for low volume fractions, the amplitude decrease is linear with $r$, which indicates that the attenuation remains small over the whole gap width, i.e. $\alpha e \ll 1$ in Eq.~(\ref{eq:prms}), and that acoustic propagation can be considered to be in the single-scattering regime. In this respect, PS-1 particles exhibit almost no  attenuation since the amplitude profile $\bar{P}(r)$ remains almost flat up to $\phi_0 =0.4$. For other types of particles, the slope of the $\bar{P}(r)$ profiles significantly increases with the global volume fraction above $\phi_0 \simeq 0.1$ for PMMA and glass spheres and as soon as $\phi_0 \gtrsim 0.03$ for PS-2 and PA spheres. This indicates that the effective attenuation of the suspension increases with $\phi_0$ but linearity implies that attenuation remains constant across the gap. Finally, for the latter two types of particles, $\bar{P}(r)$ strongly deviates from a linear profile at the largest volume fractions. Such exponentially decreasing profiles are characteristic of multiple scattering~\cite{Page:2000}.

Based on the above features of the ultrasonic amplitude, we may now proceed with calibrating the volume fraction from the $\bar{P}(r)$ profiles for which single scattering is a good approximation. For all suspensions, these profiles are fitted over their central region $0.3 \lesssim  r \lesssim 1.7$ as $\bar{P}(r)=P_0 \exp[-\alpha (e-r)]$, where the adjustable parameters $P_0$ and $\alpha$ both depend on $\phi_0$. This approximation corresponds to Eq.~(\ref{eq:prms}) with a space-independent attenuation coefficient $\alpha=1/\bar{P}\, {\rm d} \bar{P}/{\rm d}r$. In other words, thanks to the small gap of our Couette cell,  the $r$-dependence in Eq.~(\ref{eq:atten}) can be neglected so that the attenuation coefficient is a function of the volume fraction only.

\subsubsection{Calibration in the ``ideal'' case of PS-1 particles in water}

\begin{figure}
    \includegraphics[scale=0.95]{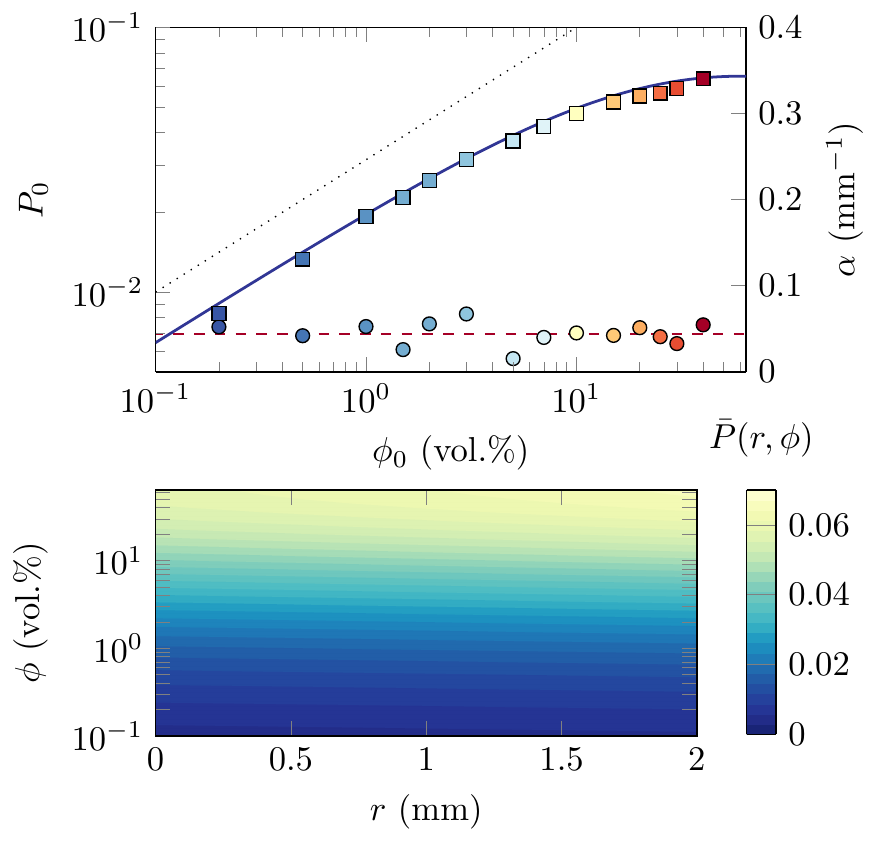}
    \caption{Calibration parameters for a suspension of PS-1 spheres in water inferred from linear fits of the ultrasonic amplitude profiles shown in Fig.~\ref{fig:C_r}. Top panel: Maximum acoustic amplitude $P_0$ (squares, left axis) and attenuation coefficient $\alpha$ (circles, right axis) as a function of the volume fraction $\phi_0$. Same color scale as in Fig.~\ref{fig:C_r}. Black dots show power-law behavior with exponent 0.5. The blue solid line is the best fit of $P_0$ using Eq.~(\ref{eq:fitfit1}). The red dashes show the average value of $\alpha$. See Supplemental Table~1 for parameter values. Bottom panel: ultrasonic amplitude $\bar{P}(r,\phi)$ obtained through the calibration process. The local volume fraction $\phi(r,z)$ can then be recovered from any measured $P(r,z)$ by inverting the $\phi$--$\bar{P}$ relationship at the corresponding radial position $r$.}
    \label{fig:calibcalib}
\end{figure}

The top panel of Fig.~\ref{fig:calibcalib} shows the fit parameters $P_0$ and $\alpha$ found for PS-1 spheres suspended in water as a function of the global volume fraction $\phi_0$. The small values of $\alpha\simeq 0.05$~mm$^{-1}$ (i.e. $\alpha e\simeq 10^{-1}$) confirm that the attenuation is negligible in this system up to surprisingly large concentrations $\phi_0\simeq 0.4$. Moreover, the scaling $P_0\sim \phi_0^{1/2}$ holds over more than a decade at low volume fractions. As shown by the red line in Fig.~\ref{fig:calibcalib}, $P_0$ is well fitted over the whole range of volume fractions by the following empirical form:
\begin{equation}
	\label{eq:fitfit1}
P_{0_{\rm fit}}(\phi) = p_0\, \frac{\phi^{1/2}}{1 + \mu \phi^{\xi}}\,,
\end{equation}
where $p_0$ and $\mu$ are two positive constants and the exponent $\xi=0.85$ accounts for the deviation from the simple square-root scaling for volume fractions larger than a few percent.

From Eq.~(\ref{eq:fitfit1}) and by taking a constant value for $\alpha$, defined as the average of all fitted values (see red dashes in Fig.~\ref{fig:calibcalib}), we may now interpolate the local ultrasonic amplitude $\bar{P}(r,\phi)$ for any value of $r\in [0,e]$ and $\phi\in [0,0.45]$ by using $\bar{P}(r,\phi)=P_{0_{\rm fit}}(\phi) \exp[-\alpha (e-r)]$. The corresponding $\bar{P}(r,\phi)$ map is shown in the bottom panel of Fig.~\ref{fig:calibcalib}. Since this equation provides a bijective relation between $P_0$ and $\phi$, it is straightforward to convert any measured ultrasonic amplitude $P(r,z)$ into a local volume fraction $\phi(r,z)$ through:
\begin{equation}
	\label{eq:invers}
\phi(r,z) = P_{0_{\rm fit}}^{-1}\left[ P(r,z) \exp[\alpha (e-r)] \right]\,.
\end{equation}
In practice, the inverse function $P_{0_{\rm fit}}^{-1}$ is solved for numerically.

\subsubsection{Calibration in the general case}

\begin{figure}
    \includegraphics[scale=0.95]{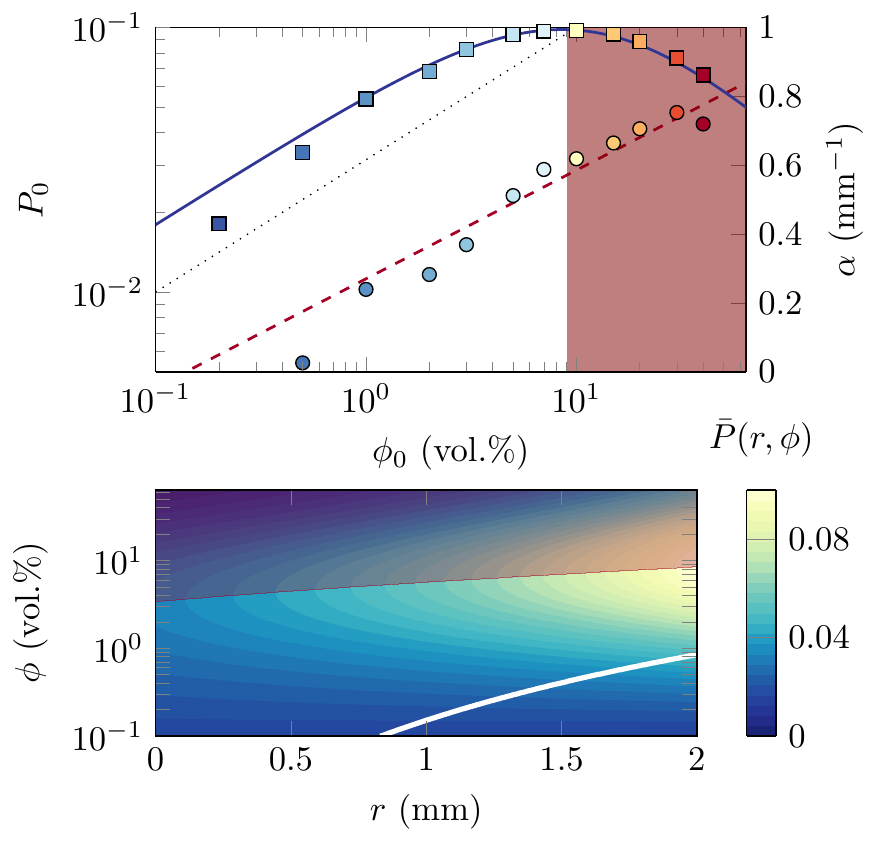}
    \caption{Same as Fig.~\ref{fig:calibcalib} for PS-2 spheres suspended in water. The red dashed line is the best fit of $\alpha$ using Eq.~(\ref{eq:fitfit2}). See Supplemental Table~1 for parameter values. In the bottom panel, the shaded area above the red line highlights the region of the $(r, \phi)$ plane where $\left . \partial \bar{P} / \partial \phi \right |_r$ is negative. Above the white line, $\bar{P}(r,\phi)$ at fixed $r$ is no longer single-valued.}
    \label{fig:calib2}
\end{figure}

The previous suspension of PS-1 particles in water constitutes an ideal case where the attenuation is small and independent of $\phi$. In general, however, $\alpha$ increases with $\phi$ such that amplitude profiles may cross each other. As the volume fraction is increased, this occurs for larger and larger ranges of values of $r$ starting from the rotor side where the effect of attenuation on $\bar{P}$ is the largest. In other words, there is no longer a one-to-one correspondence between $\bar{P}(r)$ and $\phi$. Rather, for a given radial position $r$, $\bar{P}(r)$ may go through a maximum with $\phi$. At large volume fractions, this precludes the simple inversion used above in Eq.~(\ref{eq:invers}).

Nevertheless, we found that Eq.~(\ref{eq:fitfit1}) still provides good fits of the parameter $P_0$ as a function of the volume fraction for all types of particles. We also observed that the increase of the attenuation coefficient with volume fraction could be reasonably modelled as:
\begin{equation}
	\label{eq:fitfit2}
\alpha_{\rm fit}(\phi) = \alpha_0 + \alpha_1 \log\phi + \alpha_2 \phi\,,
\end{equation}
with either $\alpha_1=0$ (which corresponds to a linear dependence of the Beer-Lambert type) or $\alpha_2=0$ (which corresponds to a logarithmic dependence). The case where $\alpha_0\ll 1$ and both $\alpha_1=0$ and $\alpha_2=0$ corresponds to the ideal case found above for PS-1 particles. Fitting the parameters $P_0$ and $\alpha$ extracted from the amplitude profiles respectively to Eqs.~(\ref{eq:fitfit1}) and (\ref{eq:fitfit2}) allows us to get a satisfactory analytical description of the $\bar{P}(r)$ data set as a function of $\phi_0$ for all particle types. We then apply the same procedure as in the ``ideal'' case but we restrict the analysis to the range of $r$ and $\phi$ for which $\phi$ can be unambiguously recovered from $P(r,z)$ as illustrated in Supplemental Figure~2.

The case of PS-2 spheres is shown in Fig.~\ref{fig:calib2} while calibration parameters for PMMA, glass and PA particles are displayed in Supplemental Figure~3. All fit parameters are gathered in Supplemental Table~1. In the case of PS-2 spheres, we find that the attenuation increases logarithmically with volume fraction as already reported in dense suspensions of glass beads~\cite{Stolojanu:2001}. However, due multiple scattering, the amplitude profiles quickly get non-linear when the concentration is increased and, as soon as $\phi_0\gtrsim 0.09$, the fits of the $\bar{P}(r)$ data become questionable as highlighted in red in the top panel of Fig.~\ref{fig:calib2}). Moreover, for all radial positions $r_0$, $\bar{P}(r_0,\phi)$ exhibits a maximum at $\phi_{\rm max}(r_0)$ (see red line in the bottom panel of Fig.~\ref{fig:calib2}): one gets $\phi_{\rm max}\simeq 0.04$ close to the rotor and $\phi_{\rm max}\simeq 0.09$ close to the stator. Volume fractions lying beyond $\phi_{\rm max}(r_0)$ are shaded with red in the bottom panel of Fig.~\ref{fig:calib2}. This implies that the $\bar{P}(r)$--$\phi$ relationship is no longer single-valued and, for the range of volume fractions investigated here, the $\bar{P}(r)$--$\phi$ relationship remains unambiguous only below the white line shown in the bottom panel of Fig.~\ref{fig:calib2}. This strongly restricts the application domain of our calibration and values outside this domain will be tagged as ``Not a Number'' (NaN) values in the following. Finally, we note that the cases of PMMA and glass particles are intermediate between those of PS-1 and PS-2 spheres while PA particles show even more dramatic non-linearity than PS-2 spheres (see Supplemental Figure~3 ).

\subsection{Potential improvements of the technique}
\label{sec:physlim}

As explained above, the most limiting factor that prevents a direct, quantitative correspondence between $\bar{P}$ and $\phi$ is the strong attenuation and non-linearity induced by multiple scattering. From a general point of view, scattering is more intense for large particles and for large acoustic impedance mismatch between the particles and the suspending fluid. Therefore, it is not surprising that the smallest particles (PS-1) and a limited impedance mismatch lead to less attenuation and give access to larger volume fractions than other suspensions made of bigger particles with larger impedance mismatch (see Table~\ref{tab:props}).

Multiple scattering is also obviously promoted by high particle volume fractions for which the ultrasonic mean free path is reduced due to the large number of interfaces between fluid and particles. In the strong scattering regime where $\bar{P}(r)$ profiles get exponential, the ultrasonic propagation can be described as diffusive \cite{Page:1995} and theoretical approaches of the multiply scattered signal could be used in the backscattering geometry to estimate the diffusion coefficient and the transport mean free path \cite{Tourin:2000}. However, linking such quantities to the \emph{local} volume fraction seems out of reach since local information is lost when multiple echoes dominate.

In the intermediate regime where $\bar{P}(r)$ profiles remain linear but cross each other due to the increase of attenuation with volume fraction, we can ensure that multiple scattering remains limited by checking that the time- and $z$-averaged velocity profiles $\bar{v}(r)$ are linear at $\dot\gamma = 20$~s$^{-1}$. Indeed, for such a laminar flow in a homogeneous suspension, significant deviations from linearity can only occur in $\bar{v}(r)$ through a breaking of the time-space correspondence due to multiple scattering. The fact that the echo from the rotor remains clearly visible in the raw speckle images also provides another check that single scattering is dominant. In such cases, one could probably extend the present calibration method by incorporating a space-dependent attenuation coefficient $\alpha(r)$ [see e.g. Eq.~(\ref{eq:atten})] and by adopting an iterative method starting from the stator to invert the multi-valued $\bar{P}(r)$--$\phi$ relationship. Such an involved inversion method is, however, outside the scope of the present paper.

\section{Case studies}
\label{sec:Examples}

Based on the calibration method described in the previous section, we now turn to the investigation of suspensions where the particle distribution is spatially heterogeneous. We focus here successively on particle migration in laminar, unstable and turbulent Taylor-Couette flows, on sedimentation under shear, and on shear-induced resuspension.

\subsection{Migration effects in cylindrical Couette flows}

The cylindrical Couette flow generated by the present rheo-ultrasonic device provides a prototypical flow to test our technique for measuring local volume fraction. Indeed, shear flows of Newtonian fluid in TC geometry are well-known to display a classical sequence of various flow regimes from laminar flow to steady vortices to turbulence as the rotation speed of the rotor is increased \cite{Taylor:1923,Andereck1986,Tritton:1988}. The transitions between the various flow regimes are controlled by the Taylor number that quantifies the ratio of (destabilizing) inertia to (stabilizing) viscosity: ${\rm Ta} = \dot{\gamma}^2 e^5 / \nu^2 R_1$, where $\nu$ is the kinematic viscosity of the fluid. 

A suspension of solid particles in a Newtonian fluid can still be considered as a Newtonian fluid as long as it is dilute enough enough, typically for volume fractions smaller than 5~\% \cite{Russel1992}. Therefore it follows the same flow sequence with the Taylor number. However, when the particles are not neutrally buoyant, they undergo flow-induced migration leading to a spatially heterogeneous volume fraction. In the following paragraphs, we use our imaging technique in dilute suspensions for various TC flows.

\subsubsection{Laminar Couette flow}

When the Taylor number is below a critical value, ${\rm Ta}<{\rm Ta}_{\rm c}=1712$, the flow field ${\bf u}$ remains laminar and purely tangential ${\bf u} = u_\theta {\bf e_\theta}$. However, large, non neutrally-buoyant particles laden in the flow can migrate, and their velocity then reads ${\bf v} = v_r {\bf e}_r + u_\theta {\bf e}_\theta$. In the simple case of a laminar Couette flow, flow-induced particle migration has two origins: particle inertia and inter-particle collisions. The first effect pushes particles denser (resp. lighter) than the fluid outwards (resp. inwards), whereas the second term originates from gradients of both local shear rate $\dot \gamma$ and particle volume fraction $\phi$~\cite{Morris:1999,Dbouk:2013}.

\begin{figure}
	\centering
    \includegraphics{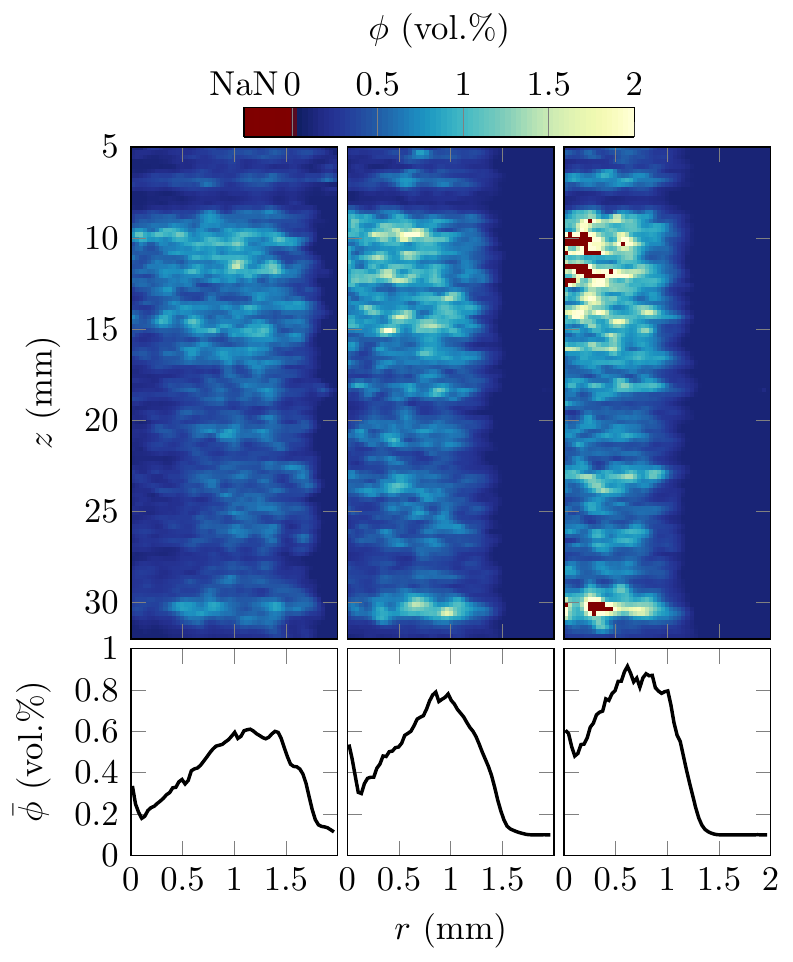}
    \caption{Radial migration of PS-2 spheres in a 27~vol.~\% water--73~vol.~\% glycerol mixture sheared at $\dot{\gamma} = 200$~s$^{-1}$, from left to right, for $t = 4$, 80 and 155~s after full mixing. Top panels:  concentration map $\phi(r,z)$. Bottom panels: $z$-average volume fraction $\bar{\phi}(r)$. The global particle volume fraction is $\phi_0=0.5$~\%. For each concentration map, the ultrasonic data were time-averaged over 100 successive acquisitions performed within 0.01~s. See also Supplemental Movie~1.}
    \label{fig:centrif}
\end{figure}

Migration is evidenced in Fig.~\ref{fig:centrif} and Supplemental Movie~1 for PS-2 particles in a 27~vol.~\% water--73~vol.~\% glycerol mixture. A depleted region is clearly seen to expand from the stator as a function of time. The front velocity is about 3.5~$\mu$\mps. Using a simple balance between particle inertia and viscous drag, the radial particle velocity is given by $v_r \simeq 2 a^2 \Delta \rho u_\theta^2 / 9 R_1 \eta $, where $\eta$ is the dynamic viscosity and $\Delta \rho$ the density difference between the particle and the surrounding liquid. With a mean orthoradial velocity of the order of $e\dot{\gamma}/2$, the above prediction leads to 3.7~$\mu$\mps, which is in good agreement with the experimental value. Long after the last panel of Fig.~\ref{fig:centrif}, the local particle volume fraction eventually reaches a heterogeneous steady-state fixed by a balance between migration, boundary effects and interparticle forces~\cite{Abbott1991,Mills1995}.

Additional particle concentration heterogeneities are visible along the $z$-direction. The observed general decreasing trend of $\phi(z)$ with increasing $z$ is expected since the particles are less dense than the suspending fluid and should ultimately cream to the top of the TC cell. The slight apparent vertical stratification is however more surprising as the Taylor number for $\dot\gamma = 200$~s$^{-1}$ in the water--glycerol mixture considered here is ${\rm Ta}\simeq 400$, which should be low enough to exclude any flow instability. Longer experiments under similar conditions have ultimately led to the formation of ``ring'' patterns, whereas imposing $\dot\gamma = 20$~s$^{-1}$ did not show any stratification. The origin of this phenomenon would probably deserve more attention in future work.

Finally, by averaging over the $z$-direction, we can check that the estimated local volume fraction is consistent with the global amount of particles $\phi_0= 0.5~\%$, at least for $t=4$ and $80$~s (see bottom panels of Fig.~\ref{fig:centrif}). For longer times, patches where the local volume fraction is not accessible lead to an underestimation of the $z$-averaged $\phi$ but the migration front is still clearly detectable.

\subsubsection{Steady Taylor vortex flow}
\label{sec:tvf}

\begin{figure}
	\centering
    \includegraphics{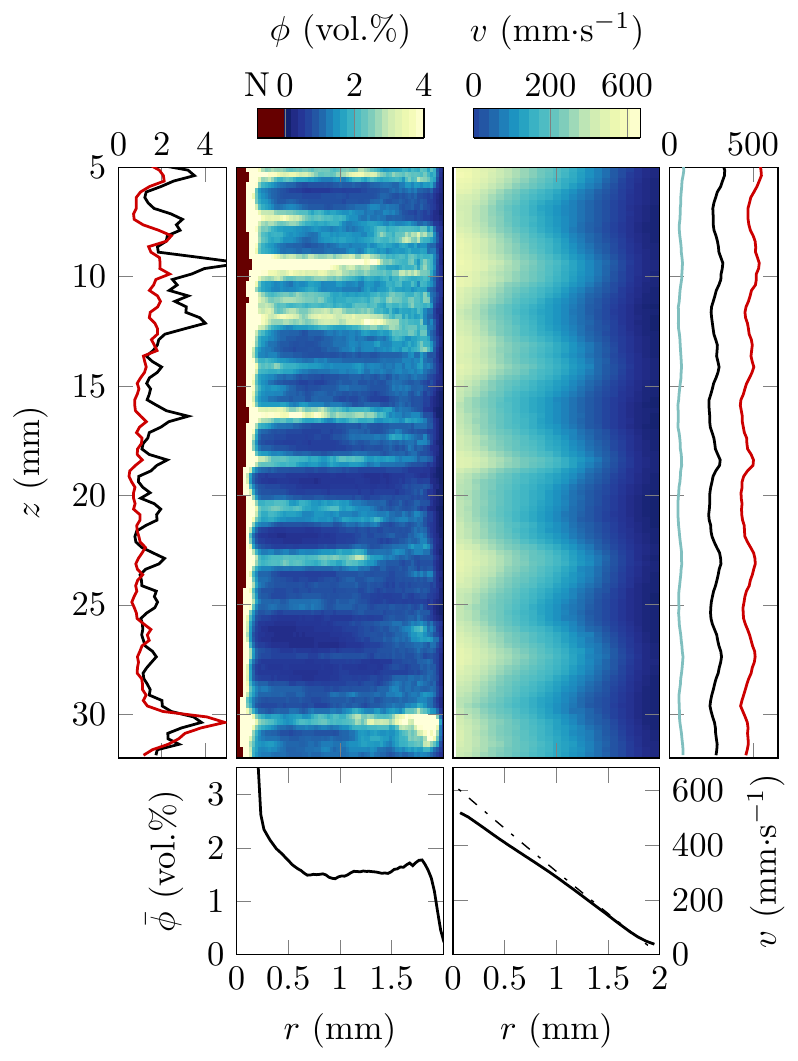}
    \caption{Rheo-ultrasonic imaging of the Taylor vortex flow of a dilute suspension of PMMA spheres in a 38~vol.~\% water--57~vol.~\% glycerol--6~vol.~\% CsCl mixture sheared at $\dot{\gamma} = 305$~s$^{-1}$. 
    Left image: volume fraction map $\phi(r,z)$ (dark red and ``N'' in the color bar corresponds to ``Not a Number'' outputs from the calibration procedure) together with its local averages for $r\simeq 0.2$~mm (left, in red) and $r\simeq 1.8$~mm (left, in black) and the $z$-average $\bar{\phi(r)}$ (bottom). 
   Right image: velocity field $v(r,z)$ together with its local averages for $r\simeq 0.2$~mm (right, in light blue), all $r$ values (right, in black) and $r\simeq 1.8$~mm (right, in red) and the $z$-average $\bar{v}(r)$ (bottom).
  The global particle volume fraction is $\phi_0=2~\%$. The ultrasonic data were time-averaged over 10,500 successive acquisitions performed within 1.8~s.} 
    \label{fig:TVF}
\end{figure}

When the Taylor number is increased above ${\rm Ta}_{\rm c}$, the laminar flow gives way to the Taylor vortex flow (TVF)~\cite{Taylor:1923,Andereck1986} where pairs of counter-rotating vortices stacked along the vertical direction superimpose to the base tangential flow. In the presence of particles, this unstable yet steady flow induces a complex migration pattern. Figure~\ref{fig:TVF} shows the combined volume fraction and velocity measurements performed at steady state in a dilute suspension of PMMA particles in a water--glycerol mixture in the presence of cesium chloride (CsCl) for a Taylor number ${\rm Ta}\gtrsim {\rm Ta}_{\rm c}$. The velocity field displays oscillations in the $z$-direction that are typical of TVF as already reported elsewhere~\cite{Gallot2013,Fardin:2014a}. In particular, the oscillation period is about 4~mm which corresponds to the expected distance $2e$ between two pairs of vortices.

Moreover, in the present case, where particles (with density $\rho_{\rm p}=1200$~kg\,m$^{-3}$) are lighter than the suspending fluid (with density $\rho_{\rm f}\approx 1300$~kg\,m$^{-3}$), we expect particle migration towards the center of the Taylor vortices promoted by inertial centripetal forces due to the steady vortex rotation. The volume fraction map $\phi(r,z)$ shown in Fig.~\ref{fig:TVF} indeed exhibits a heterogeneous particle distribution in the presence of TVF: stratification along the $z$-direction is clearly visible with a characteristic spacing of about 2~mm between more concentrated bands. The $z$-average volume fraction is consistent with the global volume fraction $\phi_0=2~\%$. However, the maxima in particle volume fraction do not match the centers of the vortices (i.e. the nodes of the velocity field along $z$) but rather coincide to the edges of the vortices (i.e. to the extrema of the velocity field along $z$). 

In the extreme case of bubbles suspended in a TVF ($\rho_{\rm p} / \rho_{\rm f} \ll 1$), experiments and simulations have shown migration towards the centers of the vortices leading to the formation of characteristic horizontal bubble strings~\cite{Djeridi1999,Climent:2007}. Numerical simulations of suspensions with $\rho_{\rm p} / \rho_{\rm f}\gtrsim 1$ also imply that particles fall in retention zones located around the centers of the vortices~\cite{Wereley:1999,Henderson:2010}. The current lack of experimental or numerical results for $\rho_{\rm p} / \rho_{\rm f}\lesssim 1$ and for  particle size relative to the gap similar to ours make it difficult to draw a definite conclusion on the observed migration towards vortex edges. 

\subsubsection{Turbulent Taylor vortex flow}

\begin{figure}
	\centering
    \includegraphics{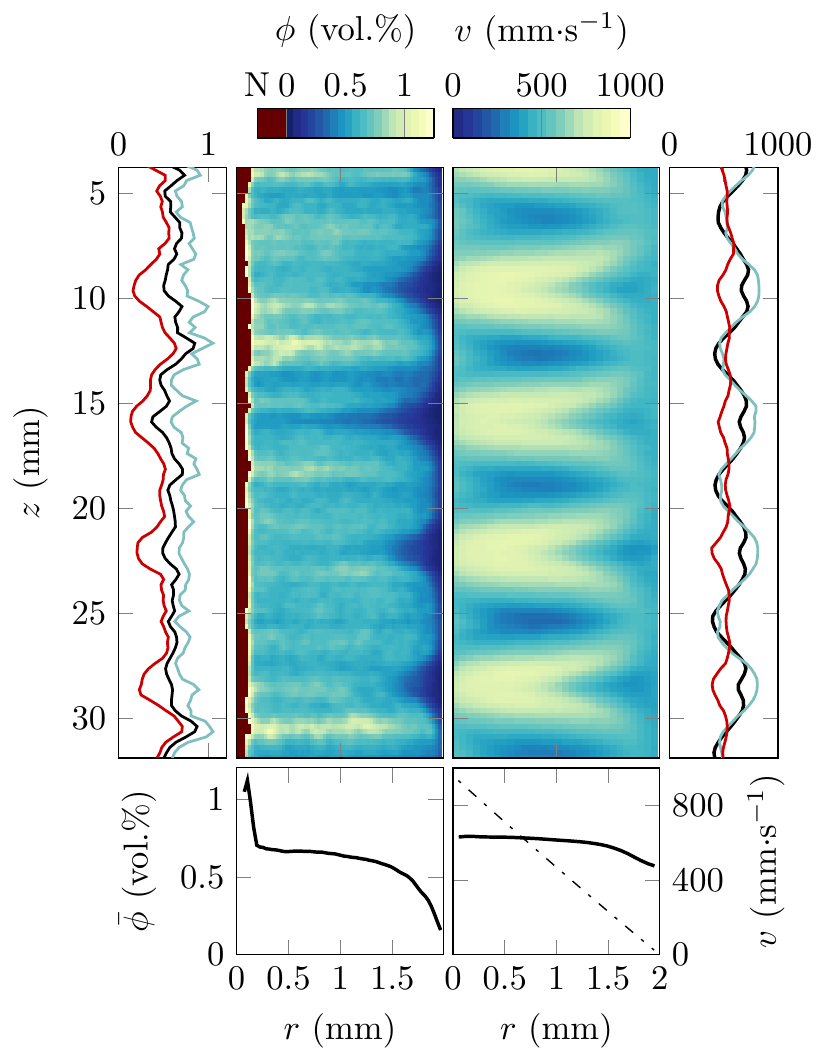}
    \caption{Same as Fig.~\ref{fig:TVF} for the turbulent Taylor vortex flow of a dilute suspension of PS-2 spheres in water sheared at $\dot{\gamma} = 500$~s$^{-1}$. The global particle volume fraction is $\phi_0=0.5\%$. The ultrasonic data were time-averaged over 5,000 successive acquisitions performed within 0.25~s.}
    \label{fig:TTVF}
\end{figure}

For very large Taylor numbers, a turbulent Taylor vortex flow (TTVF) state is reached where pairs of vortices exist only in a statistical sense with a wavelength that differs from $2e$~\cite{Andereck1986}. The turbulent instantaneous velocity field is characterized by strong velocity gradients and fluctuations so that it is difficult to neglect any term in the momentum balance of one particle in the flow~\cite{Elghobashi:1994}. Integrating the equations of motion therefore requires a very detailed knowledge of the instantaneous flow that is not accessible using our experimental technique. Conversely, particles may affect the Taylor vortex flow, as reported for volume fractions $\phi_0$ down to $1~\%$~\cite{DominguezLerma1985}. 

Figure~\ref{fig:TTVF} reports the results of rheo-ultrasonic imaging in a dilute suspension of PS-2 particles in water for a Taylor number ${\rm Ta}\simeq 3.5 \times 10^5\gg {\rm Ta}_{\rm c}$. Several turbulent vortices are well visible in the velocity field and show up in the concentration map as D-shaped zones of nearly homogeneous concentration. Outside these zones, close to the stator, the rest of the flow is almost devoid of particles. PS-2 particles are slightly denser than the surrounding fluid: in this case, applying the same arguments as in Sect.~\ref{sec:tvf} on the average flow, particles should be slowly attracted to the limit cycle surrounding the center of the Taylor vortices as described in~\cite{Henderson:2010}, which is in qualitative agreement with our observations. We also hypothesize that turbulent velocity fluctuations enhance particle diffusion and promote mixing within the vortices leading to more homogeneous concentration fields.

\subsection{Sedimentation tracking}

\begin{figure}
	\centering
    \includegraphics{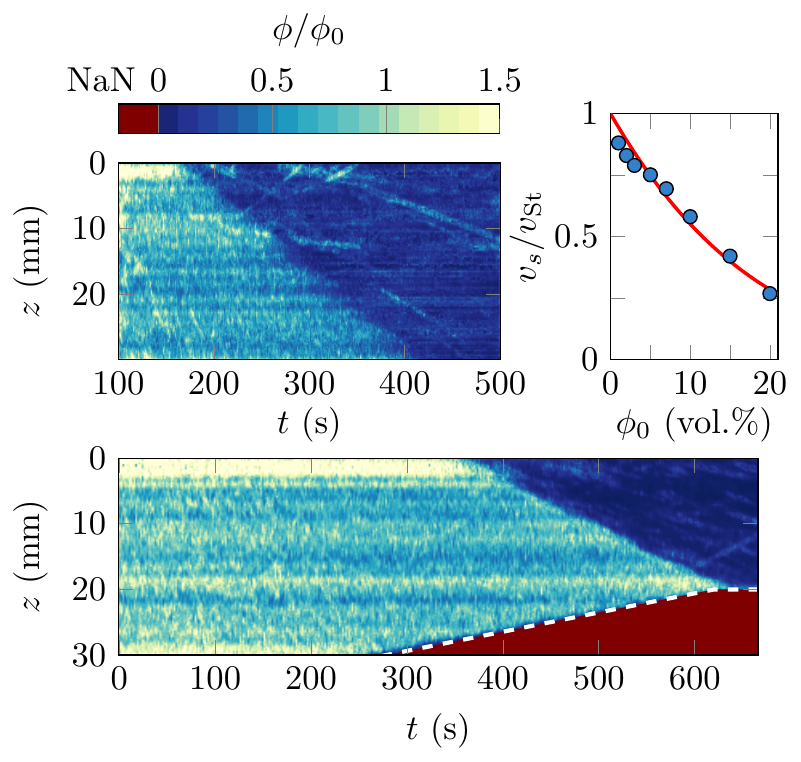}
    \caption{Sedimentation of PMMA particles in water for initially homogeneous suspensions with volume fractions $\phi_0 = 0.07$ (top) and $\phi_0 = 0.15$ (bottom). The images show spatio-temporal diagrams of the local volume fraction $\bar{\phi}(z,t)$ averaged over the radial position $r$ across the gap of the TC cell. Mixing is stopped at $t=0$ after which a zero shear rate is applied. The white dashed line in the bottom image shows the limit of the sediment that progressively grows from the bottom of the cell. For a given time $t$, ultrasonic data were averaged over 10 successive acquisitions performed within respectively 1.25 and 1.66~s. The graph in the inset shows the settling velocity $v_s$ normalized by the Stokes velocity $v_{\rm St}$ as a function of $\phi_0$ together with the theoretical prediction from Mills and Snabre~\cite{Mills1995}.}
    \label{fig:Sedim}
\end{figure}

In this second set of experiments, we use the time-resolved capacities of our technique to investigate the sedimentation of dense particles. Starting from a homogeneous suspension of volume fraction $\phi_0$, sedimentation proceeds through the creation of three layers of spatially constant concentrations: the clear fluid ($\phi = 0$), the bulk suspension ($\phi(r,z) = \phi_0$) and the sediment, considered loosely packed at some volume fraction $\phi^*$. These layers are separated by two sharp fronts where the concentration abruptly changes~\cite{Mills1994, Russel1992}. Following Ref.~\cite{Mills1994}, the velocity $v_s$ of the first front decreases with the initial volume fraction $\phi_0$ as: 
\begin{equation}
	\frac{v_{\rm s}}{v_{\rm St}} = \frac{1 - \phi_0}{1 + k \phi_0 / (1 - \phi_0)^3}\,,
    \label{eq:v_sedim}
\end{equation}
where $v_{\rm St}= 2 \Delta \rho g a^2 / 9 \nu \rho_{\rm f}$ is the Stokes velocity and the constant $k$ is set to $4.6$ to be compatible with the effective suspension viscosity determined by Batchelor~\cite{Batchelor1972}.

Figure~\ref{fig:Sedim} shows spatio-temporal diagrams of the local volume fraction $\bar{\phi}(z,t)$ averaged over $r$ as a function of vertical position $z$ and time $t$ after mixing for two initial volume fractions. These diagrams clearly highlight the dynamics of the front separating the clear fluid (where $\phi\simeq 0$, in dark  blue) from the bulk suspension. As expected, the concentration below the clear fluid-suspension front remains uniform and equal to the initial volume fraction. In addition, the experimental value of the velocity $v_s$ is in good agreement with the prediction of Eq.~(\ref{eq:v_sedim}), as already reported for experiments up to $\phi = 0.57$~\cite{Mills1994}.

Moreover, for the larger concentration $\phi_0=0.15$, the second front separating the suspension from the sediment [where $\phi(r,z)$ cannot be quantitatively estimated, in dark red in the bottom panel in Fig.~\ref{fig:Sedim}] is also visible: the sedimented particles accumulate at the bottom of the container located at $z\simeq 40$~mm and enter the field of view of the ultrasonic probe for $t\simeq 280$~s. The settling time of the suspension, which corresponds to the intersection of the two fronts, can be estimated as $t\simeq 630$~s. This time can be computed theoretically based on Ref.~\cite{Mills1994} by considering the velocity of the two concentration fronts, $v_{\rm s}$ given by Eq.~(\ref{eq:v_sedim}) and $v^*=v_{\rm s}(\phi/\phi^*)/(1-\phi/\phi^*)$ which is inferred from a mass balance. This yields a sedimentation time $t = H/(v_{\rm s} + v^*) = 590$~s that is in good agreement with the observed settling time.

\subsection{Resuspension under shear}

\begin{figure}[t]
	\centering
    \includegraphics{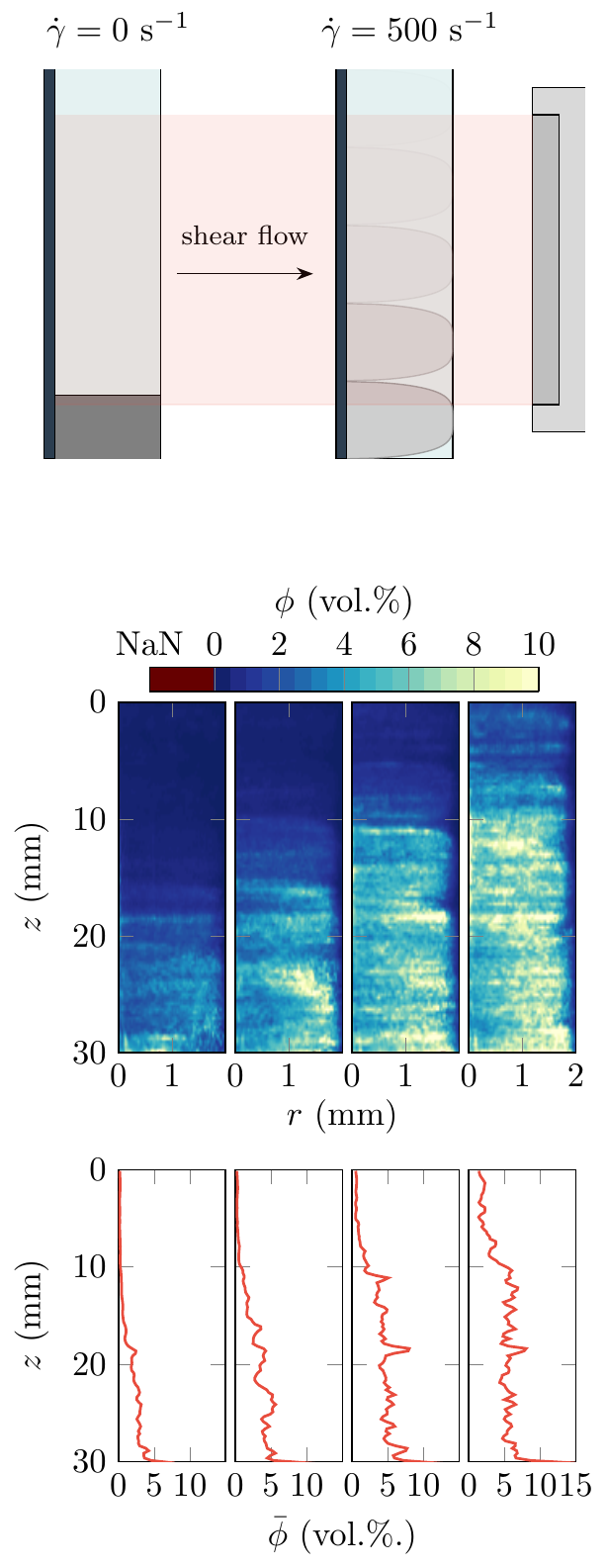}
    \caption{Shear-induced resuspension of PS-1 spheres in water with a global volume fraction $\phi_0 = 0.15$. Top panel: sketch of the experimental configuration. At rest, the sediment (in dark gray) extends from the lower end of the ultrasonic beam down to the bottom of the container. When a shear rate $\dot{\gamma} = 500$~s$^{-1}$ is applied, a turbulent Taylor vortex flow sets in and particles progressively fill the vortices (see shades of gray).
    Center panels: Concentration maps $\phi(r,z)$ for $t = 0.8$~s, $t = 1.6$~s, $t = 2.4$~s, $t = 3.6$~s from left to right. Bottom panels: corresponding $r$-averaged volume fractions $\bar{\phi}(z)$. For each concentration map, ultrasonic data were averaged over 1,000 successive acquisitions performed within 0.05~s. See also Supplemental Movie~2.}
    \label{fig:Turb_resuspension}
\end{figure}

While sedimentation or viscous resuspension by a laminar Couette flow are rather slow processes~\cite{Leighton1986,Acrivos1993}, resuspension by a turbulent flow is a much faster transient phenomenon. Figure~\ref{fig:Turb_resuspension} shows that our imaging technique can capture the resuspension of PS-1 particles by a turbulent TC flow thanks to the high repetition rate of ultrasonic pulses. 
Time-resolved concentration maps under a high shear rate $\dot{\gamma} = 500$~s$^{-1}$ display the progressive invasion of the turbulent vortices by particles (see also Supplemental Movie~2). In the steady state (similar to Fig.~\ref{fig:TTVF}), the particle concentration remains almost homogeneous in each turbulent vortex, concentration gradients being the largest at the interfaces between vortices which are more difficult to cross. Therefore, in the case of Fig.~\ref{fig:Turb_resuspension}, a homogeneous state is not reached after 3.6~s and the upper part of the cell is eventually filled with particles on longer time scales.

\section{Conclusion}

In this paper, we have demonstrated the possibility of time-resolved local particle concentration measurements under shear in a Taylor-Couette cell, based on an ultrafast rheo-ultrasonic device that provides additional simultaneous access to local velocity and standard rheological data. Quantitative concentration measurements require an extensive calibration procedure of the backscattered ultrasonic amplitude. This procedure involves an exponential approximation of the amplitude profiles $P(r)$ to define a maximum acoustic amplitude $P_0$ and an attenuation coefficient $\alpha$. These parameters are subsequently modeled analytically in order to invert the relationship between the backscattered intensity and the local particle concentration. The maximum accessible volume fraction strongly depends on the particle size and type and ranges from a few percents up to concentrations close to jamming in the most favorable cases. The linear approximation for $P(r)$ fails when multiple scattering is present and/or when the attenuation becomes too high, resulting in non-linear acoustic intensity profiles for concentrated suspensions and in sediments. 

Our device was used to assess various heterogeneous concentration fields of particle suspensions in Newtonian fluids. We have shown that shear-induced migration along the radial direction could be characterized in a laminar Couette flow, as well as non-trivial particle distributions in both steady and turbulent Taylor vortex flows. Thanks to the high temporal resolution of plane wave ultrasonic imaging, sedimentation and shear-induced resuspension processes can also be studied on time-scales down to 10~ms. Further applications of the technique include the investigation of opaque suspensions of industrial interest, e.g. for food processing and waste water management. From a fundamental point of view, it will also be applied to the understanding of particle migration in viscoelastic suspending fluids under shear, for instance in the presence of elastic or shear-banding instabilities.

\section*{Acknowledgments}
This work was funded by the Institut Universitaire de France and by the European Research Council under the European Union's Seventh Framework Program (FP7/2007-2013) / ERC grant agreement No.~258803.

\bibliographystyle{apsrev4-1}
\bibliography{biblio}

\end{document}